# Positioning Crenarchaeal tRNA-Introns

Smarajit Das [a], Zhumur Ghosh [a], Jayprokas Chakrabarti [a, b, *],
Bibekanand Mallick [a] and Satyabrata Sahoo [a]

(a) Computational Biology Group (CBG)

    Theory Department

    Indian Association for the Cultivation of Science

    Calcutta 700032 India

(b) Biogyan

    BF 286, Salt Lake

    Calcutta 700064 India

**\* Author for correspondence**

    Telephone: +91-33-24734971, ext. 281(Off.)

    Fax: +91-33-24732805

    E-mail: tpjc@iacs.res.in ; biogyan@vsnl.net

**Abstract**

**Background**: Introns play a key role in deciphering tRNA genes .Their positions have to be fixed consistent with the conserved sequence motifs in tRNA . The bulge-helix-bulge secondary structures at the exon-intron boundaries need careful attention .

**Results**: We precisely position a noncanonical intron in the odd second copy of tRNA$^{Asp}$(GTC) gene in the newly sequenced crenarchaea *Sulfolobus acidocaldarius*. The uniform assortment of some features from normal aspartate tDNA and some from those corresponding to unnatural amino acids conduce us to conjecture it to be a novel tRNA gene – probably coding for a modified aspartate residue. Further we reposition intron in tRNA$^{His}$(GUG) gene in *Pyrobaculum aerophilum*. We argue this intron repositioning enhances the structural stability of the tRNA, retaining all its archaeal-tRNA-signatures. The bulge-helix-bulge motifs at the exon-intron boundaries are re-analyzed and found to support our conjectures.

**Conclusions**: We conclude *Sulfolobus acidocaldarius* contains a new tRNA form for some nonstandard amino acid derived from aspartate . The histidine tRNA in *Pyrobaculum aerophilum* has its intron located at an altogether different position .

## Background

Archaeal tRNA genes (tDNAs) have been a subject of our recent work [1]. While working on crenarchaeal cytoplasmic tDNAs we find the need to define new introns and reposition some existing (NCBI annotated) ones to improve the secondary structure [2. In the newly sequenced crenarchaea *Sulfolobus acidocaldarius* we identify a second copy of aspartate tRNA gene with curiously odd attributes. We define a noncanonical intron (NCI) in its acceptor (A-arm). The interpretation of this tRNA is delicately balanced. It could be the aspartate tRNA gene or it could correspond to some unnatural amino acid. Sifting through the evidences and comparing with the pyrrolysine tRNA we opt for the nonstandard amino acid choice. In addition to this, the present study establishes an in-silico approach to reposition intron in the histidine tDNA in *P. aerophilum* and argues in favor of this new intron position over the old one.

Archaeal tDNA introns are located mostly in anticodon (AC) loop between tRNA positions 37/38. These are the canonical introns (CI). Archaeal tDNAs also have introns at positions other than canonical. These noncanonical introns in archaeal tDNAs were observed in 1987 [3]. Presence of introns in tDNAs at canonical position is detected by the algorithm tRNAscan-SE [4]. But detecting the noncanonical introns could be difficult. This is due to the lack of prior knowledge regarding their lengths and exact locations. A structural deviation in the secondary structure of the tRNA from the standard one hints towards improper positioning of the intron. For instance, in the NCBI-annotated crenarchaeal genome of *P. aerophilum*, tRNA$^{Gly}$(CCC) gene ranges between 663679 and 663772, with a 19 base long CI. But Marck and Grosjean records the above-mentioned sequence range to be tRNA$^{Met}$(CAU) gene with a 20 base long NCI between 29/30 [5]. Although the sequence range carries signatures of both these possibilities, the BHB structural motif appears to favour the second possibility over the first. Again, the conflict between tRNA$^{Ser}$(GCT) and tRNA$^{Met}$(CAU) has been addressed by Marck and Grosjean for this species. These amusing anomalies led us to re-investigate crenarchaeal genomes for precise positioning of the introns.

Archaeal splicing endonucleases recognize and cleave a conserved secondary structure motif called the bulge-helix-bulge (BHB) that appears at the intron-exon boundaries. Archaeal splicing machinery cleaves introns at variable positions in pre-tRNAs within the BHB motif [6]. Splicing of introns hence is a RNA-protein interaction which requires mutual recognition of complementary tertiary structures. The conformation of the BHB motif is more important for archaeal endonuclease recognition than its sequence [7].

From the time when the triplet nature of the genetic code was revealed there have been speculations that some of the sixty-four codons encode a few of the rare amino acids found in proteins [8]. The great majority of the nonstandard amino acids are created by chemical modifications of standard amino acids. Some of these may be a post translational modification, but a few are directly specified by the genetic code as in the case of selenocysteine (TGA) [9, 10] or pyrrolysine (TAG). *Methanosarcina barkeri* genome revealed the presence of pyrrolysine tRNA gene. Both the selenocysteine and pyrrolysine systems use tRNAs that are first charged by standard amino acids (serine and lysine respectively) and are then modified into non-standard derivatives (selenocysteine and pyrrolysine). Selenocystiene pathway is widespread in all kingdoms of life. But pyrrolysine is restricted to certain archaea and eubacteria [11]. The NCI defined odd-second-copy of tRNA$^{Asp}$ has interesting structural similarities with pyrrolysine tRNA.

For the present work all tDNAs of crenarchaeal genomes were investigated. Our conclusions agreed with what is known in all but two cases. In the odd-second-copy of pre-tRNA$^{Asp}$ of *Sulfolobus acidocaldarius* DSM 639 (NC_007181), we define a new NCI in A-arm. This unnatural tRNA is speculated to perform the function of coding modified form of Aspartate residue. In pre-tRNA$^{His}$ gene of *Pyrobaculum aerophilum* (NC_003364) we propose new splicing position for the intron.

Results and Discussions: Positioning Introns

*Sulfolobus acidocaldarius*

**Structural peculiarities of the odd tRNA$^{Asp}$(GTC) gene:** This gene lies between (427581..427656) . We define an NCI of length 6 lying between 3/4. Amongst the identity elements of archaeal aspartate tRNA gene [1], G34, U35, C36 and G6:C67 are present in this tRNA. But curiously A73 instead of G73. Amongst the conserved base pairs G26:U44 and U54:A58 are present. But again curiously G8:U14 rather than U8:A14 of normal Asp tRNA. Deviations also include the absence of the 16$^{th}$ and the 17$^{th}$ nucleotides and remarkably also of the normally conserved G19. Additionally in D-loop 15C and 18C instead of 15G and 18G. The BHB of the newly-defined NCI in A-arm has h$^{//}$ bh$^{/}$ L motif (h$^{//}$ is the helix formed by base pairing between the sequence atctt upstream of the tRNA gene and the exonic sequence tgggg between 12-16 of the unspliced tRNA). The splice sites are marked by arrows in fig 1a.

Note that NCBI annotated gene between (427588-427653) has an anomalous tRNA-secondary structure with just 3 base-pairs in A-arm. There is further an extra base again in A-arm.

**Similarities with tRNA pyrrolysine**: The *M. barkeri* genome has a tRNA$^{Pyl}$ gene [10] coding for the 22$^{nd}$ amino acid Pyrrolysine. Some of the secondary structural features of it match with this odd aspartate tRNA gene of *S. acidocaldarius*. In both the tRNAs the 16$^{th}$, the 17$^{th}$ and the 19th bases are absent. U35 and N36 (here N does not include U) are important identifying features of pyrrolysine tRNAs. Remarkably these are there in this odd aspartate tRNA. The characteristic peculiarities of this odd tRNA$^{Asp}$ and its resemblance with tRNA$^{Pyl}$ of *M. barkeri* are marked in fig 2.

**New Intron Splice-Site in *P.aerophilum* :**

**tRNA$^{His}$ gene** : Using our tRNA-gene identification routine we identified it between c(21869..21775). We define an NCI of 16 bases lying at 43/44. The conserved C50:G64, G29:C41 of archaeal tRNA$^{His}$ are consistently there in this tRNA. The key element for proper aminoacylation C73, unique to tRNA$^{His}$ , is also present here. G26:A44 and G31:C39 too match well with all other crenarchaeal histidinyl tRNAs [1]. In addition for the newly defined NCI we have checked the hallmark of a proper intron - the BHB. At the exon-intron boundary we find h$_e$bh/L motif with splicing positions marked by arrows in fig 1b. The 4bp long exonic helix (h$_e$) is followed by a bulge (b). In this bulge lies the 3$^/$ as well as the 5$^/$ splice sites. It is followed by another helix (h$^/$), followed by a loop (L) [h$^/$ is sometimes purely intronic. Again, intron-exon portions sometimes shuffled together within the same helix].

Note the same region of the genome is annotated in NCBI to be histidine tRNA gene with a canonical intron of 13 bases. Now, with this definition of canonical intron, 31:39 base pair is absent. This is very unusual. Again the short V-arm is composed of four to five bases in all tDNAs. The exceptions are tDNA$^{Ser}$ and tDNA$^{Leu}$ (these have long V-arms) [12, 5]. Importantly, G26:G44 and the three extra bases between 43 and 49 distorts the tRNA secondary structure from that of other crenarchaeal tRNA$^{His}$(GUG).

Precise positioning of the NCI, as done here, removes the extra bases between 43 and 49, generates proper base pairing at 31-39 and at 26-44. These improve the structure of the resulting tRNA.

.

## Conclusions:

The great majority of the nonstandard amino acids are created by chemical modifications of standard amino acids. These may be post translational modifications, or else directly specified by the genetic code as is the case with selenocysteine (TGA) or pyrrolysine (TAG). Selenocysteine and pyrrolysine tRNAs are first charged with serine and lysine, chemically modified thereafter [8, 10]. This implies that these tRNAs contain certain features of normal tRNAs as well as some special features for the modification of the attached amino acid. The secondary features of the odd aspartate tRNA gene of *S. acidocaldarius* have certain features of archaeal tRNA$^{Asp}$ gene. G34, U35 and C36 along with G73 are the major identity elements of aspartate tRNA [1, 13]. All the identity elements except the base at 73$^{rd}$ are present in this tRNA$^{Asp}$. The normal purine base G73 is replaced by another purine A73 for this odd one. In addition to these deviations from the normal aspartate tRNA, certain key similarities with pyrrolysine (corresponding to modified lysine residue) lead us to the ansatz that it is a tRNA for an unnatural amino acid , perhaps , of a modified aspartate residue.

Further we re-positioned intron in tRNA$^{His}$ from *P. aerophilum.* Precise fixing of intron splice site restructure the anticodon base pairing including the 3D- one between 26:44. The hallmark of intron-exon boundaries is the presence of a sequence capable of folding into the characteristic bulge-helix-bulge (BHB). A consistent BHB in this establishes our intron positioning.

**Methodology : Bioinformatic Input**

The tRNA search programs tRNAscan-SE [4] and ARAGORN key on primary sequence patterns and/or secondary structures specific to tRNAs. But a few loopholes exist .This has to do with the inability of these existing algorithms to identify tRNA genes if it harbours noncanonical introns, specially if these are in A-stem, T-loop and AC-stem. Although unusually located introns in tDNA were observed in 1987 [3], insilico identification of tDNAs harbouring these continues to be a challenge. Some tRNAs are misidentified or missed altogether. We studied [1] about one thousand archaeal tRNAs. From the data we developed and fine-tuned our procedure to annotate CI as well as NCI containing tRNA genes from genomes. The salient features of our procedure to locate introns are the followings: (i) Introns are considered anywhere in tRNA gene, but required to harbour the BHB motif for splicing out during tRNA maturation. (ii) After the introns are spliced out the resulting tRNA has to generate the regular cloverleaf structure. (iii) There are some exclusive crenarchaeal tRNA specific identity features. The features considered in the search are G7:C66, U55, U54:A58, G52:C62 and G53:C61. These are considered as conserved for all crenarchaea.(iv) The constraints of lengths of stems of A-arm, D arm, AC arm and T arm are 7,3, 5 and 5 base pair respectively.

(v) Base positions optionally occupied in D-loop are 16,17, 17a, 19, 20a and 20b, (vi) an extra arm or V arm is considered for type II tRNAs; its length constrained to maximum of 21 bases.

Once we find a candidate tRNA we sift through our data [1] to position the introns with proper BHB at the intron-exon boundaries. This has to be done consistent with all conserved sequence features. Identification of tRNA depends on these conserved bases and base-pairs.

### Authors' contributions

SD and ZG analyzed the BHB motifs at the exon-intron boundaries ; SD,ZG,JC and BM conceived and implemented the main ideas presented here . SS developed the necessary softwares . The work was drafted by ZG and JC .

### Acknowledgements :

We acknowledge useful discussions and critical comments from Dhananjay Bhattacharyya.

Intron 3/4

Figure 1a

Intron 43/44

Figure 1b

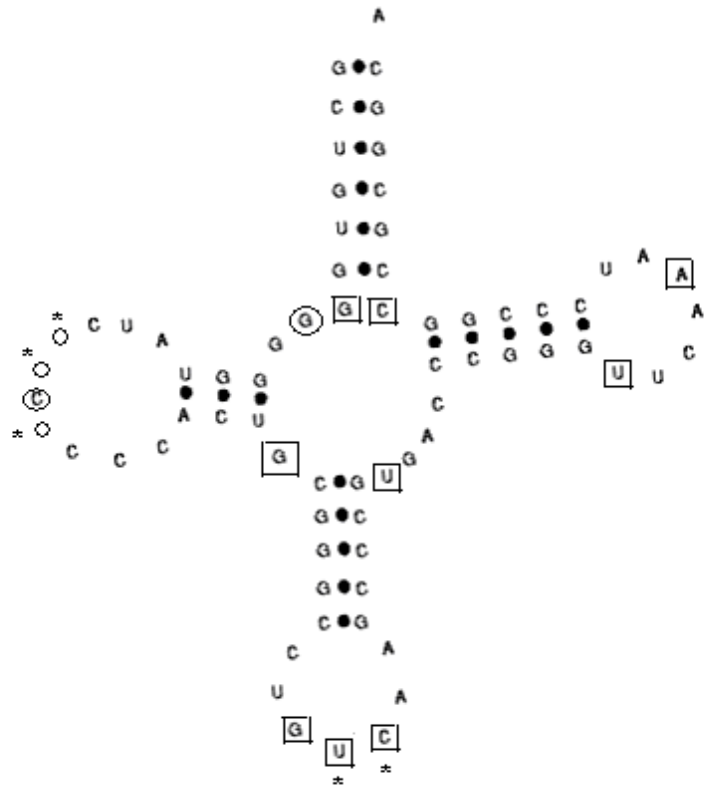

**Figure 2**

**Figure Legends**

**Figure 1**:

NCIs denotes start position of noncanonical intron.

NCIe denotes end position of noncanonical intron.

h$^{/}$ is sometimes purely intronic and sometimes intron-exon portion shuffled together within the same helix.

Figure 1a: Unusual tRNA$^{Asp}$ with the BHB structure of the intron in the A-arm.

h$^{//}$ is the helix formed by base pairing between the sequence atctt upstream of the tRNA gene and the exonic sequence tgggg between 12-16 of the unspliced tRNA.

b denotes bulge. h$^{/}$ here is purely intronic.

Bases 2-15 of the unspliced tRNA are replaced within conventional cloverleaf scheme by asterisks and are drawn in an alternative 2D structure that fits the splicing motif (h$^{//}$bh/L) requirement.

Figure 1b: tRNA$^{His}$ with the BHB structure of the intron between 43/44.

h$_e$ denotes exonic helix. b denotes bulge. h$^{/}$ here is purely intronic.

Bases 40-52 are replaced within conventional cloverleaf scheme by asterisks and are drawn in an alternative 2D structure that fits the splicing motif (h$_e$bh/L) requirement.

**Figure 2**: Cloverleaf structure of the odd tRNA$^{Asp}$ showing similarity with tRNA$^{Pyl}$ and deviation from normal archaeal tRNA structure.

The conserved bases of aspartate tRNA are boxed. The empty circles and circled bases are non-canonical (those which deviates from universal features) archaeal tRNA features. The asterisk marked bases match with pyrrolysine tRNA of *M. barkeri*.